\documentclass[11pt]{article}
\usepackage{moriond,epsfig}

\def\be{\begin{equation}}
\def\ee{\end{equation}}
\def\bea{\begin{eqnarray}}
\def\eea{\end{eqnarray}}

\def\la{\mathrel{\hbox{\rlap{\hbox{\lower4pt\hbox{$\sim$}}}\hbox{$<$}}}}
\def\ga{\mathrel{\hbox{\rlap{\hbox{\lower4pt\hbox{$\sim$}}}\hbox{$>$}}}}
\def\amin{^\prime}

\def\asec{^{\prime \prime}}
\begin{document}
\vspace*{4cm}
\title{FIRST RESULTS FROM THE XMM-NEWTON OBSERVATION IN THE
       GROTH-WESTPHAL SURVEY STRIP REGION}

\author{Takamitsu Miyaji, Richard E. Griffiths}

\address{Department of Physics, Carnegie Mellon University\\
         5000 Forbes Av., Pittsburgh PA 15213, USA}

\maketitle\abstracts{
 We present the first results of the XMM-Newton observation
on the Groth-Westphal Strip Field. The scheduled exposure time 
was 70 [ks]. We first analyzed $\sim 50$ [ks] of calibrated and cleaned 
EPIC-PN data, which were the only ones we could successfully process
with an early version of the SAS software applicable to these observations. 
A total of $\sim 110$ sources have been detected in the central 10 arcmin 
of the PN field of view (FOV). A cross-correlation with the Hubble
Space Telescope (HST) Medium Deep Survey (MDS) 
database shows a range of morphological properties of host galaxies of 
the X-ray source. Spectra of hard X-ray sources in our data indeed show 
absorption in the range $N_{\rm H}\sim 10^{22}-10^{23} {\rm [cm^{-2}]}$
at $z\sim 1$. }

\section{Introduction}

  The Groth-Strip field is one of the best-studied
fields in extragalactic astronomy. The Groth-Strip Campaign 
started from 28 contiguous medium-deep HST
Wide-Field Planetary Camera 2 (WFPC2) images 
by E. Groth\cite{groth}, followed by numerous follow-up projects
over the entire electromagnetic spectrum. 

 In this field, we are investigating the population of X-ray sources 
using our XMM-Newton guaranteed time observation (PI=Griffiths), with an
emphasis on morphological properties of the  X-ray sources, taking 
advantage of the sharp WFPC2 images and the HST MDS 
database\cite{mds} with extensive disk-bulge morphological 
characterizations.

\section{The Data}

 The observations were made during revolutions 111-114 (July 2000).
The period of low particle background is concentrated in revolution 113. 
We were able to generate a cleaned and calibrated dataset for only 
EPIC-PN from the available software at the time of this preliminary analysis. 
With the most recent version of SAS (5.1 beta) at the time of writing
this paper, we have been able to generate calibrated 
datasets for the MOS detectors also and we are starting
the analysis of the combined PN and MOS dataset.

\begin{figure}

\psfig{file=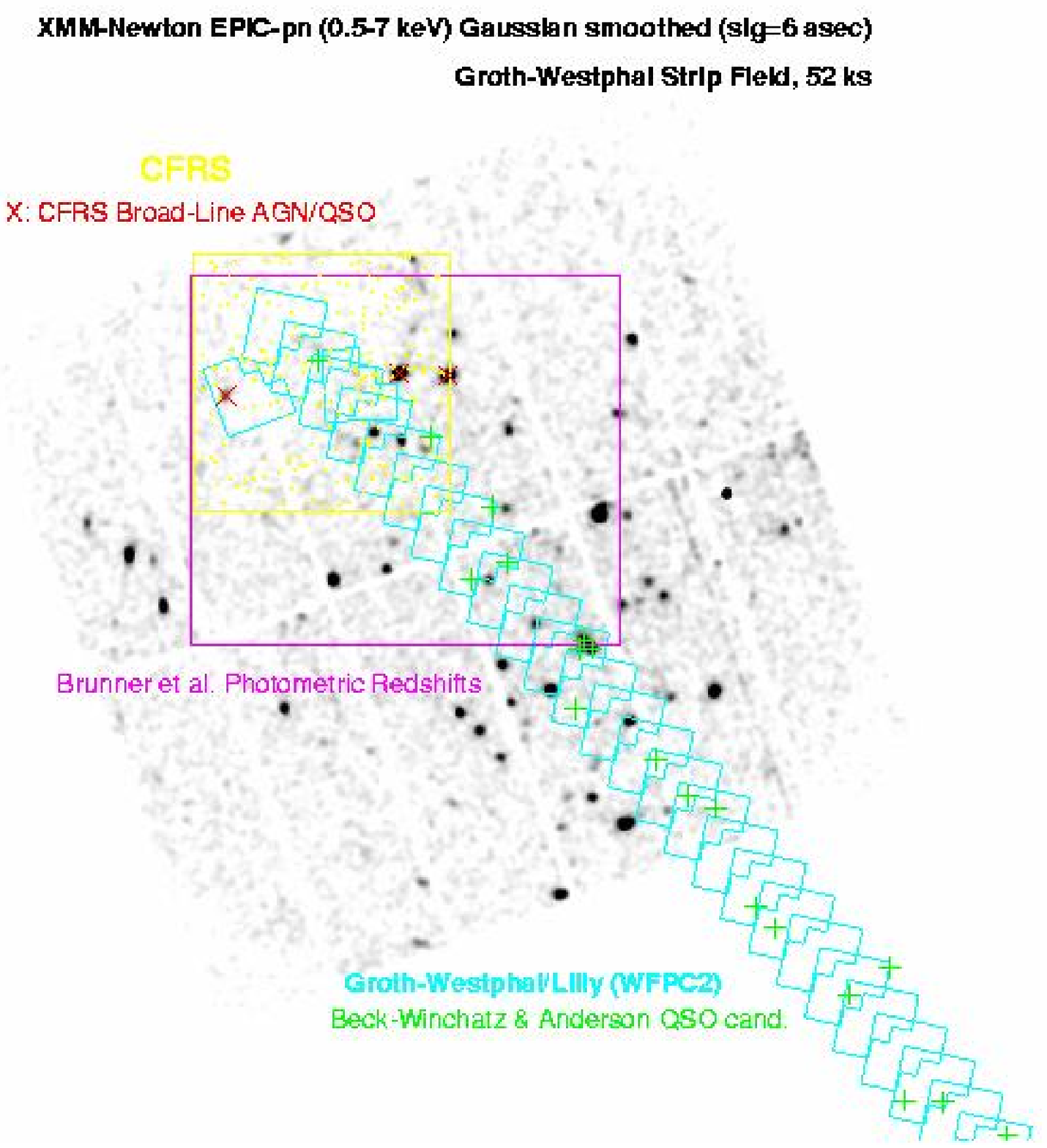,width=0.9\hsize,clip=}
\caption[]{\raggedright 
The XMM-Newton EPIC-PN image (underlying grayscale) of the
Groth-Strip, smoothed with a $\sigma=6\asec$ gaussian, is
shown with WFPC2 FOVs (cyan), the CFRS\cite{cfrs} FOV and
galaxies with redshifts (yellow), broad-line AGNs in the CFRS (red cross) 
and the region with the photometric redshift catalog by 
Brunner~et~al.\cite{bru} (magenta). The XMM-Newton PN FOV is 
$27\amin \times 27\amin$.\label{fig:groth_field}}
\end{figure}

 Three broad-line AGNs in the CFRS v1.0\cite{cfrs} catalog and 
three AGN candidates in  Beck-Winschatz \& Anderson~\cite{bwa} (BWA)
have X-ray counterparts in our EPIC-PN image. Positions of the X-ray
sources have been determined by single or multiple-source  elliptical 
gaussian fits. The X-ray data have been aligned by matching these
positions with known optical positions of these AGNs. 
The positional errors of the fits (90\% along each axis) were
$\la 1\asec$. The residuals of the alignment were typically  
$\la 1\asec$ and the largest residual was $2\asec$. 
 Fig. \ref{fig:groth_field} shows the PN image (smoothed with 
a $\sigma=6\asec$ gaussian) with a number of overlays as 
indicated in the caption.

\section{Preliminary Source Detection}
 
 Preliminary source detection has been made on the 50 ks
of EPIC-PN data using the automated source-detection task 
{\em edetectproc} in SAS 5.0 with a likelihood threshold of 
$-\ln P = $ 15, where $P$ is the probability that the source
does not exist (approximately corresponding to a $5\sigma$ 
detection). The numbers of sources in different bands ($N_{\rm det}$) 
and  the approximate detection thresholds ($S_{\rm x}^{\rm min}$) are 
shown in Table~\ref{tab:srcs}. There are unremoved hot pixels 
and columns in our current data and the source detection 
procedure also recognized peaks of these features as 
real sources. Table~\ref{tab:srcs} shows the number of the sources 
which apparently come from the hot columns also ($N_{\rm hot}$). 
Note that we are likely to miss the recognition of spurious sources 
which are more subtle.  Exclusion of all spurious sources will 
be possible in the future by matching up the PN and MOS source lists.
An approximate detection limit for each band is shown. For the
count rate-to flux conversion, the PIMMS software was used with 
a $\Gamma=1.4$  power-law with Galactic absorption 
($N_{\rm H}=1\cdot 10^{20}{\rm [cm^{-2}]}$).  

\begin{table}
\caption{Groth-Westphal Field EPIC-PN 50 ks} 
\label{tab:srcs}
\begin{center}
\begin{tabular}{|ccc|}
\hline
%\multicolumn{3}{|c|}{324 arcmin$^2$} \\
Energy & $N_{\rm det}$ ($N_{\rm hot}$)$^{\rm a)}$ & 
    $S_{\rm x}^{\rm min\;\;\;b)}$  \\
 ${\rm [keV]}$  &             & ${\rm [erg\,s^{-1}\,cm^{-2}]}$ \\ 
\hline
0.5-7.0 & 113 ($\ga 6$) & $2 \cdot 10^{-15}$ \\
0.5-2.0 &  92 ($\ga 9$) & $7 \cdot 10^{-16}$ \\
2.0-4.5 &  39 ($\ga 0$) & $2 \cdot 10^{-15}$ \\
4.5-10  &  18 ($\ga 0$) & $8 \cdot 10^{-15}$ \\
\hline
\end{tabular}
\end{center}
$^{\rm a)}$ Number of sources over 324 arcmin$^2$. $^{\rm a)}$ 
Sensitivity varies over the FOV and the approximate detection
limits show here are estimated near the optical axis. 

\end{table}

\section{X-ray sources in the HST WFPC2 Fields}

 In the strip of the HST WFPC2 fields, BWA\cite{bwa} selected 
20 QSO/AGN candidates based on the existence of point-like cores 
and UV-excess down to ($B\la$ 24.5), among which 11 objects are 
within our EPIC-PN FOV. A quick match with our PN data immediately 
revealed the difficulty of selecting AGNs from numerous objects in 
the optical/UV data. Out of the 11 AGN candidates and 12 X-ray sources 
in the overlapping area, only 3 sources match. While there are 
possibilities that some of the BWA AGN candidates are actually 
X-ray weak AGNs (e.g. BAL QSOs), it is also possible that their 
compact-core UV-excess criteria selects galaxies with circum-nuclear 
starbursts as well.

\begin{figure}
\begin{center}
\psfig{file=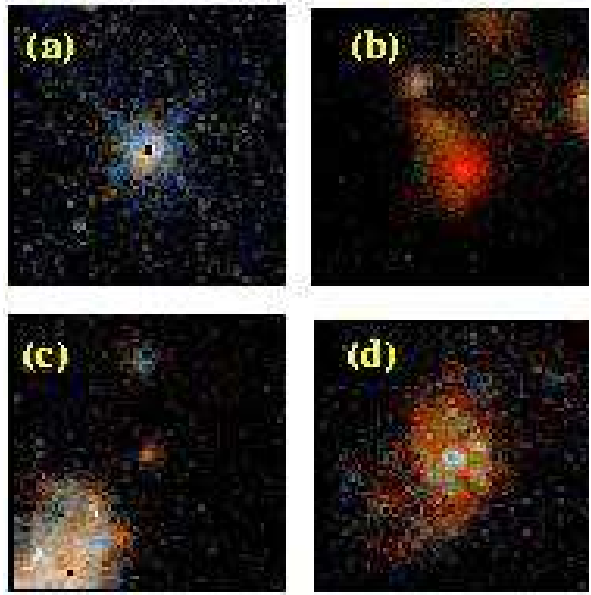,width=0.5\hsize,angle=270,clip=}
\end{center}
\caption[]{\raggedright 
The WFPC2 morphologies of selected X-ray sources in the
Groth-Strip Filed are shown. The images have 64$\times$ 64 pixels with
0.1$\asec$/pixel(a)(c)(d) or 0.046$\asec$/pixel (b).
(a): a broad-line QSO with z=1.6, stellar image (b): Seyfert 2,
z=1.15, the X-ray spectrum shows absorption and the WFPC2 image
can be fitted with a pure-bulge. Possibly interacting.
(c) the X-ray source is a very faint galaxy (not a point source) with
F606W$=26.2$ [mag]. (d) $z\sim$ 0.8 Seyfert 1. Disk+Bulge WFPC2 image.
\label{fig:morph}}
\end{figure}

 The existence of HST WFPC2 images in this field immediately enables us 
to investigate the morphological properties of the X-ray sources. In 
particular, it allowed immediate comparison with the MDS\cite{mds} database, 
which contains the results of extensive disk-bulge separations for galaxies.
Table~\ref{tab:morph} shows the number of X-ray sources in each 
morphological category in the MDS database (for the F814W filter, 
where the stellar population in host galaxies is more enhanced
when compared with the star-like AGN). The WFPC2 images (from the 
MDS database\cite{mds}) for four of the X-ray sources are shown
in Fig. \ref{fig:morph} with known properties in the caption.

\begin{table}
\caption{MDS Morphologies of X-ray sources }
\label{tab:morph}
\begin{center} 
\begin{tabular}{|ccccc|}
\hline
Stellar & Disk+Bulge & Pure Bulge & Galaxy$^{\rm a)}$ & No MDS entry \\
   3    &      2     &     3      &     2         &      2       \\
\hline
\end{tabular}
\end{center}

$^{\rm a)}$ An extended component has been resolved but the galaxy is 
  too faint for a disk/bulge discrimination.
\end{table}

\section{Spectra of Selected Sources}

\begin{figure}
\begin{center}
\psfig{file=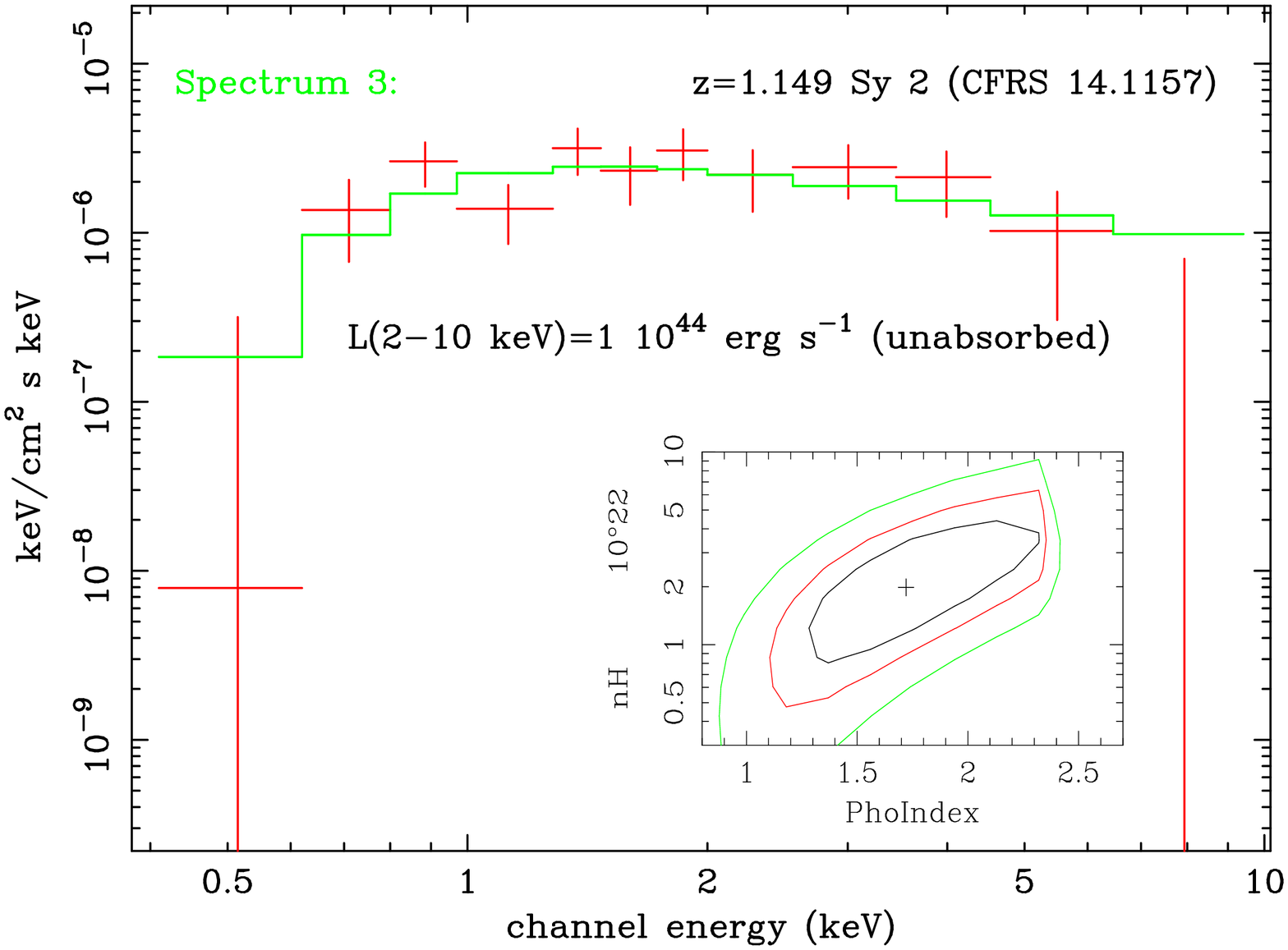,width=0.35\hsize,angle=270,clip=}
\psfig{file=src2_spec.eps,width=0.35\hsize,angle=270,clip=}
\end{center}
\caption[]{\raggedright
EPIC-PN unfolded Spectra of two of hard X-ray sources
on the Groth-Westphal Strip region are shown. Note that the
absolute flux calibration of the response matrix used in these 
fits was still highly uncertain.\label{fig:spec}}
\end{figure}

 XMM-Newton enabled us to investigate spectra of faint 
($S_{\rm x}\la 10^{-14}$ ${\rm [erg\,s^{-1}\,cm^{-2}]}$)
sources. With {\it ASCA}, {\it Beppo-SAX} and {\it Chandra}
data, it was not still clear whether these faint ``hard'' 
sources have actually absorbed spectra, intrinsically hard, 
and/or reflection-dominated. Fig. \ref{fig:spec} shows two
examples of absorbed spectra found in the XMM-Newton data.
These clearly indicate absorbed spectra.

\section{Summary}

\begin{enumerate}
\item First results from the XMM-Newton observations on the Groth-Westphal
  Strip Filed have been presented. About 50 [ks] of EPIC-PN data have
  been analyzed thus far.  
\item With a preliminary source detection, we obtained $\sim 110$ X-ray
  sources to a  limiting flux of $2\times 10^{-15}$ 
  ${\rm [erg\,s^{-1}\,cm^{-2}]}$ (0.5-2 [keV]) in the central 10 
  arcmin-radius region, a small fraction of which are still unremoved 
  hot pixels and columns.  
\item We have investigated the morphological properties of the 
  optical counterparts of the 12 X-ray sources which are within 
  the HST WFPC2 FOVs. The X-ray sources show a variety of morphological 
  properties.
\item The large effective area of XMM-Newton enabled us to obtain
  good spectroscopic data on faint hard X-ray sources. They indeed show
  evidence that they are absorbed AGNs. 
\end{enumerate}

\section*{Acknowledgments}

The results presented here are based on observations obtained with
XMM-Newton, an ESA science mission with instruments and
contributions directly funded by ESA member states and  the USA (NASA).
The EPIC instrument was developed by the EPIC Consortium led by the 
Principal Investigator Dr. M. J. L. Turner of the University of
Leicester. The data analysis presented here has been supported by 
NASA Grant NAG5-3651 to REG (XMM-Newton Mission Scientist support) 
and partially by the NASA Grant NAG5-10875 to TM (Long-Term Space 
Astrophysics).

\end{document}